\documentclass[showpacs,preprintnumbers,amsmath,amssymb,amsfots]{revtex4}

\usepackage{amsmath,amsfonts,amssymb}
\usepackage{graphicx}
\usepackage{bm}
\usepackage{mathrsfs,epigraph}

\newcommand{\lag}{\ensuremath{\mathscr{L}}}        

\newcommand{\der}{\mathscr{D}_{\bm{\xi} a}}
\newcommand{\cri}{\mathscr{C}_{\bm \xi }{}}
\def\ten{\psi_{(\ell)}}
\def\tensor#1#2{\psi^{#1}_{#2}}

\newcommand{\lie}{\ensuremath{\mathcal{L}}_{\bm{\xi}}} 

\newcommand{\beq}{\begin{equation}}
\newcommand{\eeq}{\end{equation}}
\newcommand{\beqa}{\begin{eqnarray}}
\newcommand{\eeqa}{\end{eqnarray}}
\def\nn{\nonumber\\}
\def\eq#1{(\ref{#1})}

\def\F#1#2{\ensuremath{F_{#1#2}}}           
\def\cd#1{\ensuremath{\nabla_{#1}}}          
\def\pd#1{\ensuremath{\partial_{#1}}}        
\def\tm#1#2{\ensuremath{T_{\mathscr{M}}^{#1#2}}} 
\def\tc#1#2{\ensuremath{T_{\mathscr{C}}^{#1#2}}}  
\def\tb#1#2{\ensuremath{T_{\mathscr{B}}^{#1#2}}}  

\def\t#1#2{\ensuremath{{T^{#1#2}}}}
\def\jota#1{\ensuremath{{\mathcal{J}_{\bm \xi}^{#1}}}}

\def\til#1{\widetilde{#1}}
\def\dlag#1{\frac{\partial\lag}{\partial\ \! #1}}
\def\conm#1#2{(\cd #1 \cd #2 - \cd #2 \cd #1 )}
\def\pun{{}^{\dots}_{\ \ \dots}}
\def\dv{\sqrt{| g |}\ d^n x}
\def\th#1#2#3{\widehat{T}^{#1#2#3}}
\def\st{space-time}
\def\med{\frac{1}{2}}

\begin{document}


\title{{On the  energy-momentum tensor }}

\author{Ricardo E.\ Gamboa~Sarav\'{\i}}

 \email{quique@venus.fisica.unlp.edu.ar}
\affiliation{Departamento de F\'\i sica, Universidad Nacional de
La Plata\\ C.C. 67, 1900 La Plata, Argentina
}%

\date{September 10, 2002}

\begin{abstract}
We clarify the relation among  canonical,  metric  and
Belinfante's energy-momentum tensors for general tensor field
theories. For any tensor field $\bm T$, we define a new tensor
field $\til {\bm T}$, in terms of which the Belinfante tensor is
readily computed. We show that the latter is the one that arises
naturally from Noether Theorem for an arbitrary \st\ and it
coincides on-shell with the metric one.

\end{abstract}

\maketitle

\epigraph{\em Symmetry as wide or as narrow as you may define its
meaning, is one idea which man through the ages has tried to
comprehend and create order, beauty, and perfection.}{ Hermann
Weyl \cite{Weyl}}

\section{Introduction}
For many decades a suitable definition for the  energy-momentum
tensor has been under investigation. This is more than merely  
a technical point, not only because $T^{ab}$ should  provide
meaningful physical conserved quantities but also because it is
the source of Einstein's gravitational field equations.

  In flat \st\  the canonical energy-momentum tensor arises
from  Noe\-ther's Theorem by considering the conserved currents
associated to  translation invariance. However,  only for scalar
fields the energy-momentum tensor constructed in this way turns
out to be symmetric. Moreover, for Maxwell's theory, it breaks
gauge symmetry. Of course, it is possible to correct it through
Belinfante's symmetrization procedure \cite{bel}, although this is
usually presented as an {\em ad hoc}  prescription (see for
example \cite{land,jack}).

On the other hand, a completely different approach, based on the
diffeomorphism invariance of the theory, leads to the metric
energy-momentum tensor (see for example \cite{haw}) which is, by
definition, symmetric and gauge invariant.

The aim of this paper is to clarify the relation among these
tensors.

In section 2, we define the tensor $\til {\bm T}$, which turns out
to be a very useful tool for the rest of our work. In section 3,
we analyze the relation among the different energy-momentum
tensors for general tensor field theories on an arbitrary \st\ of
any dimension.

\section{Lie derivatives and the tensor $\til {\bm T}$}
Let $\bm \xi$ be a vector field  on a (semi)Rimannian manifold of
dimension $n$ and $\phi_{t}$ a local one-parameter group of
diffeomorphism generated by $\bm \xi$. This diffeomorphism maps
 each tensor field $\bm T$ at  $p$ of the type $(r,s)$ into
 $\phi_{t*}\bm T|_{\phi(p)}$, the pullback  of  $\bm T$.

 The Lie derivative $\lie {\bm T}$ of a tensor field
${\bm T}$ with respect to $\bm \xi$ is defined to be minus the
derivative with respect to $t$ of this family of tensor fields,
evaluated at $t=0$, i.e. \beq \lie \bm
T=\lim_{t\rightarrow0}\frac{1}{t}(\bm T|_p-\phi_{t*}\bm T|_p)\ .
\eeq Thus, it measures how much the tensor field  ${\bm T}(x^a)$
deviates  from being formally invariant under the infinitesimal
transformation $x^{\prime a}= x^a -t \xi^a$, with $t\ll 1$.

 The coordinate components
are \beqa\label{lie} \lie\ T^{b_1\dots b_p}_{\ \ \ \ \ \ c_1\dots
c_q}= \pd a T^{b_1\dots b_p}_{\ \ \ \ \ \ c_1\dots c_q}\ \xi^a \nn
-\ T^{ab_2\dots b_p}_{\ \ \ \ \ \ \ c_1\dots c_q}\ \partial_a
\xi^{b_1} -\ T^{b_1a\dots b_p}_{\ \ \ \ \ \ \ c_1\dots c_q}\
\partial_a \xi^{b_2} -\dots \nn +\ T^{b_1\dots b_p}_{\ \ \ \ \ \
ac_2\dots c_q}\ \partial_{c_1} \xi^{a} +\ T^{b_1\dots b_p}_{\ \ \
\ \ \ c_1a\dots c_q}\ \partial_{c_2} \xi^{a} + \dots\ . \eeqa Of
course, for any torsion free connection, the partial derivatives
can be replaced by covariant ones.

 The following definition will prove useful.
For each tensor field $T^{c_1\dots c_p}_{\ \ \ \ \ \ d_1\dots
d_q}$ of type $(p,q)$ we define a tensor field
 $\widetilde{T}^{c_1\dots c_p}_{\ \ \ \ \ \ d_1\dots d_q}{\!}^a_{\ b}$ of type $(p+1,q+1)$,
 such that
\beqa\label{til}
 \til T^{c_1\dots c_p}_{\ \ \ \ \ \ d_1\dots d_q}{\!}^a_{\ b}:=
T^{ac_2\dots c_p}_{\ \ \ \ \ \ \ d_1\dots d_q}\ \delta^{c_1}_b +\
T^{c_1a\dots c_p}_{\ \ \ \ \ \ \ d_1\dots d_q}\ \delta^{c_2}_b
+\dots \nn -\ T^{c_1\dots c_p}_{\ \ \ \ \ \ bd_2\dots d_q}\
\delta_{d_1}^{a} -\ T^{c_1\dots c_p}_{\ \ \ \ \ \ d_1b\dots d_q}\
\delta_{d_2}^{a} - \dots\ . \eeqa For a scalar field $\varphi$ we
define $\til \varphi=0$, since there is no index to be replaced.
So, in terms of $\til T$, \eq{lie} can be written as \beqa
\label{ld} \lie\ T^{c_1\dots c_p}_{\ \ \ \ \ \ d_1\dots d_q}= \pd
a T^{c_1\dots c_p}_{\ \ \ \ \ \ d_1\dots d_q}\ \xi^a - \til
T^{c_1\dots c_p}_{\ \ \ \ \ \ d_1\dots d_q}{\!}^a_{\ b}\ \pd a
\xi^{b}\nn =\cd a T^{b_1\dots b_p}_{\ \ \ \ \ \ c_1\dots c_q}\
\xi^a - \til T^{c_1\dots c_p}_{\ \ \ \ \ \ d_1\dots d_q}{\!}^a_{\
b}\ \cd a \xi^{b}\ , \eeqa where \cd a denotes the covariant
derivative associated to the Levi-Civita connection. In index free
notation, our definition \eq{til} reads \beq \til{\bm T}(\cd {}
\bm \xi):= \cd {\bm \xi}{\bm T}- \lie {\bm T}\ . \eeq

Some simple examples are in order. For instance, for the  tensor
$\delta^a_b$ we have \beq \til\delta^a_b{}^c_{\ d}=\delta^c_b
\delta^a_d - \delta^a_d \delta^c_b =0\ , \eeq which expresses the
fact that $\lie \delta^a_b= \cd {\bm \xi}\delta^a_b=0 $. Moreover,
for the metric tensor we have \beq \til g_{ab}{}^c_{\ d}=-g_{db}
\delta^c_a - g_{ad} \delta^c_b\  , \eeq and so \beqa\label{pp}
\lie g_{ab}= \cd {c} g_{ab}\ \xi^c- \til g_{ab}{}^c_{\ d} \cd c
\xi^d= g_{db}\ \cd a \xi^d + g_{ad}\ \cd b \xi^d = \cd a \xi_b +
\cd b \xi_a\ .\eeqa On the other hand, the well known  expression
for the derivative of the volume element \beq \label{vol}\lie
\sqrt{| g |}= \med \sqrt{| g |}\ g^{ab} \lie g_{ab}= \sqrt{| g |}\
\cd a \xi^a\ , \eeq can also be   obtained from \beq \til
\varepsilon_{a_1 a_2 \dots a_n}{}^b_{\ c}= - \varepsilon_{c a_2
\dots a_n}{\delta}^b_{ a_1}- \varepsilon_{a_1 c \dots
a_n}{\delta}^b_{a_2}-\dots=- \varepsilon_{a_1 a_2 \dots
a_n}{\delta}^b_{c}\ ,\eeq where $\varepsilon_{a_1 a_2 \dots a_n}$
is the Levi-Civita alternating symbol.

Consequently, with this notation, other classical formulas of
Ricci calculus simplify: \beqa\label{pd} \cd a T^{c_1\dots c_p}_{\
\ \ \ \ \ d_1\dots d_q}= \pd a T^{c_1\dots c_p}_{\ \ \ \ \ \
d_1\dots d_q} + \Gamma^b_{\ ca}\ \til T^{c_1\dots c_p}_{\ \ \ \ \
\ d_1\dots d_q}{\!}^c_{\ b}\ , \eeqa or \beqa\label{a5} (\cd a \cd
b - \cd b \cd a )T^{c_1\dots c_p}_{\ \ \ \ \ \ d_1\dots d_q}=
 R^d_{\ cab}\ \til T^{c_1\dots c_p}_{\ \ \ \ \ \ d_1\dots d_q}{\!}^c_{\ d}\
, \eeqa where $R^d_{\ cab}$ is the Riemann curvature tensor.

Notice that \beqa \til{\mathbf{T}\otimes\mathbf{S}}=\til{
\mathbf{T}}\otimes\mathbf{S}+ \mathbf{T}\otimes\ \til{\mathbf{S}}\
, \eeqa and \beqa\label{a7} \til{\cd e T}^{c_1\dots c_p}_{\ \ \ \
\ \ d_1\dots d_q}{\!}^a_{\ b}=
 \cd e ( \til T^{c_1\dots c_p}_{\ \ \ \ \ \ d_1\dots d_q}{\!}^a_{\ b})\
- \delta^a_e\ \cd b T^{c_1\dots c_p}_{\ \ \ \ \ \ d_1\dots d_q} \
, \eeqa because  there is an additional covariant index to be
replaced in the left-hand side.

When there is no danger of confusion, we shall suppress the
unnecessary indices and write, for instance, \eq{a7} as
\beqa\label{a8} \til{\cd e T }\pun {\!}^a_{\ b}=  \cd e \til T
\pun {\!}^a_{\ b} - \delta^a_e\ \cd b T \pun\ , \eeqa or, as in
next section, even as \beqa\label{a18} \til{\cd e T } {}^a_{\ b}=
\cd e \til T {}^a_{\ b} - \delta^a_e\ \cd b T\ . \eeqa

We shall now consider the commutator $[\lie,\nabla]$ between the
Lie derivative $\lie$ and the covariant one $\nabla$ associated to
the Levi-Civita connection. It is a map which takes each smooth
tensor field of type $(p,q)$ to a smooth $(p,q+1)$ tensor field.
In the index notation, we denote the tensor field resulting from
the action of $[\lie,\nabla]$ on $T^{b_1\dots b_p}_{\ \ \ \ \ \
c_1\dots c_q}$ by $\der T^{b_1\dots b_p}_{\ \ \ \ \ \ c_1\dots
c_q}$. On scalar fields it vanishes, for \beqa [\lie,\nabla]f=
\lie df - d\lie f=0\ \ \  \text{so}\ \ \ \der f=0\ . \eeqa
Moreover, its action on the metric tensor is very simple \beqa
\der g_{bc}=-\cd a \lie g_{bc}=-\cd a(\cd b\xi_c+\cd c\xi_b)\ .
\eeqa

Since, for any two tensors $\mathbf{T}$ and $\mathbf{S}$ \beqa
[\lie,\nabla] (\mathbf{T}\otimes\mathbf{S})=[\lie,\nabla]
\mathbf{T}\otimes\mathbf{S}+ \mathbf{T}\otimes\ [\lie,\nabla]
\mathbf{S}\ , \eeqa we have \beqa\label{leib} \der
\left(T^{b_1\dots}_{ \ \ \ \ c_1\dots} \ S^{d_1\dots }_{ \ \ \ \
e_1\dots}\right) =\der (T^{b_1\dots}_{ \ \ \ \ c_1\dots}\ ) \
S^{d_1\dots }_{ \ \ \ \ e_1\dots} +T^{b_1\dots}_{ \ \ \ \
c_1\dots} \der ( S^{d_1\dots }_{ \ \ \ \ e_1\dots})\ .\nn \eeqa We
can derive the general formula for the action of  $[\lie,\nabla]$
on an arbitrary tensor field from the Leibnitz rule (\ref{leib})
if we know  its action on scalars and one-forms (or vectors).
However, we can achieve it  easier by using the symbol $\til T
\pun$, for \beqa \der T \pun =\lie \cd a T \pun - \cd a \lie T
\pun\nn = \xi^b\ \cd b \cd a T \pun - \til{\cd a T}\pun{\!}^b_{\
c} \cd b \xi^c - \cd a (\xi^b\ \cd b T \pun - \til T \pun{\!}^b_{\
c} \cd b \xi^c)\nn =\xi^b (\cd a \cd b - \cd b \cd a) T \pun +
\til T \pun{}^b_{\ c} \cd a \cd b \xi^c)\nn = (R^c_{\ bda}\ \xi ^d
+ \cd a \cd b \xi^c)\ \til T \pun{\!}^{b}_{\ c}\ , \eeqa where we
have used \eq{a8} and \eq{a5}. By defining \beqa\label{c}
\cri^c_{\ ba} :=R^c_{\ bda}\ \xi ^d + \cd a \cd b \xi^c, \eeqa for
any tensor field $ T \pun$, we can write \beqa\label{a15} \der   T
\pun   = \cri^c_{\ ba}\ \til T \pun{\!}^b_{\ c}\ . \eeqa
 Notice that $\cri^c_{\ ba}$ is symmetric in the lower indices, for
\beqa \cri^c_{\ ba} =R^c_{\ bda}\ \xi ^d + \cd b \cd a \xi^c +
(\cd a \cd b - \cd b \cd a)\xi^c\nn =  (R^c_{\ bda}+R^c_{\ dab})\
\xi ^d + \cd b \cd a \xi^c\nn =R^c_{\ adb}\ \xi ^d + \cd b \cd a
\xi^c = \cri^c_{\ ab}\ , \eeqa where we have used    the symmetry
properties of the Riemann tensor.

Now, for the metric tensor \eq{a15} reads \beqa\label{c1} \der \
g_{bc} = - \cd a \lie\ g_{bc}= -\cri_{ cba}  - \cri_{ bca}\ .
\eeqa By index substitution, we also have \beqa
 - \cd b \lie\  g_{ca}= -\cri_{ acb}  - \cri_{ cab}\label{c2}\ ,\\
 - \cd c \lie\  g_{ab}= -\cri_{ bac}  - \cri_{ abc}\ .\label{c3}
\eeqa We add equations \eq{c1} and \eq{c2} and then subtract
equation \eq{c3}. Using the symmetry property of $\cri^c_{\ ba}$
we find \beqa\label{cri} \cri^d_{\ ab} =
 \frac{1}{2}\ g^{dc} \left( \cd a \lie g_{bc} +\cd b \lie g_{ac}
-\cd c \lie g_{ab}\right)\nn = \frac{1}{2}\ g^{dc}\  \bigl( \cd
a(\cd b\xi_c+\cd c\xi_b) + \cd b(\cd a\xi_c+\cd c\xi_a)-\cd c(\cd
a\xi_b+\cd b\xi_a)\bigr)\ . \eeqa Of course, \eq{cri} can be
readily obtained by adding to the definition \eq{c} the null term
$ 3 \cd {[a} \cd b \xi_{c]}$.

Therefore, we see that  $\cri^c_{\ ba}$ are linear combinations of
the covariant derivatives of the Lie derivative of the metric
tensor. Thus we can write \eq{a15} as \beqa \der   T \pun   =
\frac{1}{2}\  \til T \pun{\!}^{bc}\ \left( \cd a \lie g_{bc} +\cd
b \lie g_{ac} -\cd c \lie g_{ab}\right). \eeqa

Notice that the action of $[\lie,\nabla]$ on any tensor field
vanishes when $\cd a \lie g_{bc}=\cd a(\cd b\xi_c+\cd c\xi_b)=0$,
and so Lie derivative and the covariant one commute in this case.
In particular, it occurs
 when $\xi^b$ is a Killing vector, so
\beqa \cd a\lie T \pun = \lie \cd a T\pun \ ,\eeqa for any tensor
field $T\pun$ when $\xi^b$ is a Killing vector field.

\section{The energy-momentum tensor}
Let us  consider a field theory where the Lagrangian $\lag$ is a
local function of a collection of tensor fields  $\tensor
{b_1\dots b_p} {(\ell) \ \ \ c_1\dots c_q}$ defined on a
(semi)Riemannian  manifold, their first covariant derivatives $\cd
a \tensor {b_1\dots b_p} {(\ell) \ \ \ c_1\dots c_q}$, and  the
metric tensor $g_{ab}$.
 Often we shall suppress all tensor indices
and denote the fields by $\ten$.

As usual, we obtain
 the equations of motion  by requiring that the action
\beq\label{action} S=\int_\Omega \lag(\cd a\ten,\ten,g_{ab})\ \dv\
, \eeq
 be stationary under arbitrary variations of the fields $\delta
\ten$ in the
interior of any compact region $\Omega$. Thus, one obtains
\beq\label{eqm}
\cd a\left(\dlag{ \cd a \ten}\
\right)=\dlag{ \ten}\ .
\eeq

\begin{figure}[h]\begin{center}
\includegraphics[height=2.5cm]{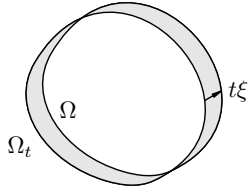}
\caption{\label{omega} The boundary contribution to equation
\eq{111}.}\end{center}
\end{figure}

The action (\ref{action}) must be  independent of the coordinates
we choose. Needles to say, that even in flat space-time we are
allowed to use curvilinear coordinates, so it must be invariant
under general coordinates transformations. By making a change of
coordinates generated by the vector field $\xi^{a}$,
$x^{a}\rightarrow x^{a}-t \xi^{a}$, the action can be written as
\beq\label{llamb} S=\int_{\Omega_t} \lag_t\ \sqrt{|g_t|}\ dx\ ,
\eeq where $\lag_t= \lag(\cd a\ \phi_{t*}(\ten),\phi_{t*}(\ten),
\phi_{t*}(g_{ab}))$, that is the same function $\lag$ evaluated on
the Lie dragged tensors fields, and $|g_t|=
\det(\phi_{t*}(g_{ab}))$. Now, taking the derivative of
(\ref{llamb}) with respect to $t$ and evaluating it at ${t=0}$ we
get three terms: \beqa\label{111} \int_{\Omega}
\frac{d}{dt}(\lag_t)_{t=0}\ \dv +\int_{\Omega} \lag\
\frac{d}{dt}\left(\sqrt{|g_t|}\right)_{t=0}\ dx
\nonumber\\+\frac{d}{dt}\left(\int_{\Omega_t} \lag\
\dv\right)_{t=0}=0\ . \eeqa The first one, by definition, contains
the Lie derivative of $\lag$; the second one,  the derivative of
the volume element  \eq{vol}, while the last one (see Fig.
\ref{omega}) is a boundary term which by using the Gauss theorem
can be rewritten as a volume integral, so we get \beqa
0=\int_\Omega \left[\lie \lag +\lag\ \cd a
\xi^{a}-\cd {a}(\lag\ \xi^{a}) \right]\dv\nonumber\\
=\int_\Omega \left[\lie \lag -\cd a(\lag)\ \xi^{a}\right]\dv\ .
\eeqa
Therefore, taking into account that the vector field $\xi^a$ is
completely arbitrary,  we have \beq\label{Lie} \lie \lag -\cd
a(\lag)\ \xi^a=0\ , \eeq which just reflects that the Lagrangian
$\lag$ must be a scalar function. Thus, the result is very simple,
the invariance of the action under general coordinate
transformations requires   $\lag$ to be an scalar function, and
(\ref{Lie}) must hold for any vector field $\xi^a$.

Now, taking into account that the Lagrangian $\lag$ depends on
the coordinates only through the tensor fields $ \cd a \ten,\
\ten$ and $g_{ab}$, we can write \eq{Lie} as
\beqa\label{7}
  \dlag{
\cd a \ten}\ \lie \cd a \ten +\dlag{ \ten }\ \lie  \ten +\dlag{
g_{ab}}\ \lie g_{ab}-\cd a(\lag)\ \xi^a=0\ , \eeqa which  is   a
linear combination of the vector field $\xi_b$ and its first
derivatives $\cd b \xi_c$.

Now, from \eq{ld} ( see also \eq{til} for our
 definition
of $\til\ten{}^c_{\ d}\ $) we can write
\beq\label{til3}
\lie \ten =
\nabla^b  \ten\ \xi_b - \til{\ten}{}^{bc} \cd b \xi_c\ ,\eeq
and, consequently
\beqa\label{til4}
\lie \cd a \ten =
\nabla^b \cd a \ten\ \xi_b - \til{\cd a \ten}{}^{bc} \cd b \xi_c\nn
= \nabla^b \cd a \ten\ \xi_b - \cd a\til{ \ten}{}^{bc} \cd b \xi_c+
\delta^b_a \nabla^c \ten \cd b \xi_c\nn
=\cd a \nabla^b  \ten\ \xi_b - \cd a\til{ \ten}{}^{bc} \cd b \xi_c+
 \nabla^b \ten \cd a \xi_b\ + (\nabla^b \cd a - \cd a \nabla^b) \ten\ \xi_b\ ,\eeqa
where in the second line we have used \eq{a8}.
Using the field
equations  (\ref{eqm}), \eq{til3} and \eq{til4} we can write the first two terms
in \eq{7} as
\beqa\label{76}
\dlag{
\cd a \ten}\ \lie \cd a \ten +\dlag{ \ten }\ \lie  \ten
=\cd a \left(\dlag{ \cd a \ten}\    \nabla^b \ten \   \xi_b
 \right) \nn
-\cd c \left(\dlag{ \cd c \ten}\    \til \ten{}^{ab} \right)  \cd
a \xi_b + \dlag{ \cd a \ten}\    (\nabla^b \cd a - \cd a \nabla^b)
\ten \   \xi_b\ . \eeqa Now, we shall rewrite the last term. From
\eq{a5} we can write \beqa\label{77}
 \dlag{ \cd a \ten}\ (\nabla^b \cd a - \cd a \nabla^b)\ten \nn = \dlag{ \cd a \ten}
 \ R^{d \ b}_{\ c\ a}\ \til{ \ten}{}_{\ d}^{c}
 =R^{b}_{\ adc}\ \dlag{ \cd a \ten}
 \  \til{ \ten}{}^{[cd]}\ ,
\eeqa
where, as usual
$ \til\ten{}^{\ [cd]}=\frac{1}{2}
(\til\ten{}^{cd}-\til\ten{}^{dc}) $.

Now,  defining
\beqa \label{th}
\th abc:=\underbrace{\dlag { \cd a \ten}\
\til\ten{}^{\ [cb]} +\overbrace{\dlag { \cd b\ten}\ \til\ten{}^{\ [ac]}
\Biggl.\Bigr.+\dlag { \cd c \ten}\ \til\ten{}^{\ [ab]} }^{\text{symmetric under\ }b\leftrightarrows c}}
_{\text{antisymmetric under\ }a\leftrightarrows b}\ , \eeqa
 we can rewrite \eq{77} as
\beq \dlag{ \cd a \ten}\ (\nabla^b \cd a - \cd a \nabla^b)\ten \
=-R^{b}_{\ adc}\ \th acd \ , \eeq since there is no contribution
from the last two terms in $\th acd$,  due to the antisymmetry of
the Riemann tensor in the last two indices. But,  using now the
symmetry properties of  $R^{b}_{\ adc}$ and $\th abc$, we get
\beqa\label{79} \dlag{ \cd a \ten}\ (\nabla^b \cd a - \cd a
\nabla^b)\ten \ = -\med\ (R^{b}_{\ adc}\ \th acd+R^{b}_{\ cda}\
\th cad)\nn = - \med\ (R^{b}_{\ acd}+R^{b}_{\ cda})\ \th cad  =
\med\ R^{b}_{\ dac}\ \th cad\nn = \med\ \conm ac \th cab= \cd a\cd
c \th cab\ , \eeqa So, we can write the last term in \eq{76} as
\beqa\label{83}
 \dlag{ \cd a \ten}\ (\nabla^b \cd a - \cd a \nabla^b)\ten\ \xi_b \
 =\cd a\cd c \th cab\ \xi_b\nn =\cd a (\cd c \th cab\ \xi_b)\ -\cd c \th cab\ \cd a\xi_b\ .
\eeqa Hence, the  first two terms in \eq{7} can be written as
\beqa\label{84}
 \cd a
\left(\dlag{ \cd a \ten}\    \nabla^b \ten \   \xi_b
+\ \cd c \th cab\   \xi_b  \right)
-\cd c \left(\dlag{ \cd c \ten}\    \til \ten{}^{ab}
+  \th cab \right)  \cd a \xi_b \ .\nn
\eeqa

Therefore, the requirement that $\lag$  be scalar leads, for any
$\xi_b$, to \beqa\label{99} \cd a \left(\dlag{ \cd a \ten}\
\nabla^b \ten \   \xi_b +\ \cd c  \th cab\  \xi_b - \lag\ \xi^a
\right) \nn +\left[2\ \dlag {g_{ab}}\ -\cd c \left(\dlag{ \cd c
\ten}\    \til \ten{}^{ab} +  \th cab \right) + g^{ab} \lag
\right] \cd a \xi_b =0 \ , \eeqa where we have used \eq{pp} and
the obvious symmetry of the tensor field $\dlag {\!g_{ab}}$.

Now, we define the canonical energy-momentum tensor as
\beqa\label{tc} \tc ab:= -\dlag{ \cd a \ten}\    \nabla^b \ten \ +
g^{ab} \lag \ . \eeqa and the metric one as \beqa\label{tm} \tm
ab:= 2\ \dlag {g_{ab}}\ -\cd c \left(\dlag{ \cd c \ten}\    \til
\ten{}^{ab} +  \th cab \right) + g^{ab} \lag . \eeqa By
definition, \tm ab\ is  symmetric,  for it follows readily from
\eq{th}, the definition of $\th abc$, that the term between
brackets in \eq{tm} is
 symmetric.
By using these definitions, for any vector field $\xi^a$, we can
write \eq{99} as \beqa\label{100} \cd a \left(-\tc ab\  \xi_b +\
\cd c \th cab\   \xi_b  \right) +\tm ab\ \cd a \xi_b =0 \ . \eeqa

 Therefore,  defining the Belinfante energy-momentum tensor
\beqa\label{tb} \tb ab:=\tc ab\ - \cd c \th cab \ = -\dlag{ \cd a
\ten}\  \nabla^b \ten \ - \cd c  \th cab  + g^{ab} \lag  , \eeqa
 we finally get
\beqa\label{110}
\cd a
\left(\tb ab\  \xi_b
 \right)
-\tm ab\ \cd a \xi_b =0 \ , \eeqa or, alternatively
\begin{eqnarray}\label{magic2}
\cd a
 \left((\tb ab - \tm ab)\ \xi_b
 \right)
+\cd a (\tm ab)\  \xi_b =0 \ .
\end{eqnarray}

Moreover, taking into account the symmetry of \tm ab, we can also write \eq{110} as
\begin{eqnarray}\label{magic}
 \cd a \left(\tb ab \xi_b \right) -\frac{1}{2}\ \tm ab\ \lie g_{ab}=0\ .
\end{eqnarray}

Equation \eq{110}, a rewritten form of \eq{Lie}, which holds for any vector field $\xi^a$,
has several important consequences. In fact,
we shall use it in five different ways:

i) Let us  restrict attention to the case where  $\xi^a$ is a
Killing vector field, i.e. a generator of an infinitesimal
isometry, so $\lie g_{ab}=\cd a \xi_b+\cd b \xi_a=0$. From
(\ref{magic}), we directly   obtain the Noether conserved current
$\jota a $ associated to this symmetry
\begin{eqnarray}\label{Noether}
\cd a \jota a = \cd a (\tb ab  \xi_b)=0
\end{eqnarray}
for, in this case, the last term in \eq{magic} clearly vanishes.
So, we can think of \tb ab as a linear function from covector
fields to vector fields such that \beqa \bm
T_{\mathscr{B}}(\text{Killing covector})= \text{conserved
current.} \eeqa

 ii) At any point of the manifold we can choose  Riemannian normal coordinates $x^\alpha$
 (i.e., a local inertial coordinate system). Moreover, we can choose for $\xi_b$ any set of $n$
linear independent covectors with constant components in this coordinate system. For instance,
the dual basis covectors $dx^{\alpha}_b$. So, in this local coordinate system,  \eq{magic}
reads
\begin{eqnarray}
 \pd \alpha (\tb \alpha\beta )\ \xi_\beta
 +  \tb \alpha\beta \pd \alpha \xi_\beta
 - \tm \alpha\beta\ \pd \alpha \xi_\beta=
 \pd \alpha (\tb \alpha\beta )\ \xi_\beta =0\ ,
\end{eqnarray}
because of  the vanishing of Chistoffel symbols and  partial
derivatives of $\xi_\beta$. Hence, we get $\cd \alpha \tb
\alpha\beta = \pd \alpha \tb \alpha\beta =0$. But this is a
tensor relation, then
\begin{eqnarray}
\cd a \tb ab =0\ .
\end{eqnarray}

iii) Now, we integrate (\ref{magic2}) over any compact region $\Omega$,
taking arbitrary vector fields $\xi^a$ vanishing everywhere except in  its  interior.
The first contribution may be transformed into an integral over the boundary
which vanishes as $\xi^a$ is zero there. Since the second term must
therefore be zero for arbitrary $\xi^a$, it follows that
\begin{eqnarray}\label{127}
\cd a \tm ab=0\ .
\end{eqnarray}

iv) Now, coming back to (\ref{magic2}),  we see that the
diffeomorphism invariance of the action yields not only $\cd a \tb
ab=\cd a \tm ab =0$, but also
\begin{eqnarray}\label{22}
\cd a\left(\ (\tb ab -\tm ab)\ \xi_b \right)=(\tb ab -\tm ab)\cd a
\xi_b=0\ ,
\end{eqnarray}
for any covector field $\xi_b$. Therefore, since $\cd a \xi_b$ is arbitrary,
we conclude that both tensors  coincide
\begin{eqnarray}
\tb ab =\tm ab\ .
\end{eqnarray}

Therefore, we have shown that
\begin{eqnarray}
\cd a \tb ab=0\ ,\ \ \cd a \tm ab=0\ ,\ \ \ \ \text{and}\ \ \ \ \
\tb ab =\tm ab\ ,
\end{eqnarray}
follow as a consequence of the diffeomophism invariance of the
action.

v) For any covector $\xi_b$, due to the asymmetry of $\th abc$, it
holds \beqa\label{1500} \cd a\left(\cd c \th cab\ \xi_b+\th cab\
\cd c \xi_b \right) =\cd a \cd c \left(\th cab\   \xi_b \right)\nn
= \med\ \conm ac \left(\th cab\ \xi_b \right)\nn =\frac{1}{2}
\left( R ^c_{\ dac}\ \th dab\ \xi_b + R ^a_{\ dac}\ \th cdb\
\xi_b\right) = R_{ac} \th cab \xi_b=0\ , \eeqa because of the
symmetry of the Ricci tensor $R_{ab}$. Thus, we can also write
\eq{100} as \beqa\label{101} \cd a \left(\tc ab\  \xi_b + \th cab\
\cd c \xi_b \right) -\tm ab\ \cd a \xi_b =0 \ . \eeqa The last
term in \eq{101} vanishes for any Killing vector field owing to
the symmetry of \tm ab. So, besides $\tb ab \xi_b$, we get an
other conserved current \beq\label{1501} \tc ab \xi_b + \th cab\
\cd c \xi_b \ ,\eeq which is, in general, linear in the Killing
vector field $\xi^a$ and its first covariant derivatives. Of
course, this current differs from $\tb ab \xi_b$ by  the
divergentless vector $ \cd c (\th cab \xi_b )$.

For scalar fields $\ten$ the last term in this current is absent,
since $\til \ten$ vanishes in this case, and both currents
coincide.

On the other hand, for general tensor fields,  this vanishing also
occurs if there exists a parallel Killing vector, i.e. $\cd a
\xi_b=0$. So, \beq \label{qq} \cd a (\tc ab \xi_b)= \cd a \tc ab\
\xi_b =0\ ,\ \ \text{for any parallel}\ \ \xi^b\ , \eeq thus, the
vector $\cd a \tc ab $ is orthogonal to $\xi^b$. Of course, this
occurs in flat \st , where we can always find $n$ linear
independent parallel vectors, for example the cartesian
coordinates vectors. Then, in that case, we have $\cd a \tc ab
=0$. But, as we are going to see, this is an exception. $\cd a \tc
ab \neq 0$ for curved \st.

Notice that, the conservation of the current \eq{1501} means that
\beqa\label{778}
 \cd a
\left(\dlag{ \cd a \ten}\ \lie \ten -\lag\ \xi^a\right)=0\ , \eeqa
which holds for any Killing vector $\xi^a$ and fields satisfying
the field equations \eq{eqm}. This result can also be readily
obtained from \eq{Lie} using the fact, shown in the preceding
section, that the Lie derivative with respect to a Killing vector
field and the covariant one commute.

\bigskip
Some comments are in order. We want to point out that \tb ab\ does
not depend   on Killing vectors. \tb ab depends  on the fields,
their derivatives and the metric, and $\cd a \tb ab =0 $ is always
true, even when the metric has no isometry at all. But, of course,
 a tensor by itself does not give rise to any
conserved quantity\footnote{ For $\cd a \t ab
=\dfrac{\partial_a(\sqrt{-g}T^{ab})}{\sqrt{-g}}+ T^{ac}\
\Gamma^b_{\ ca}\ .$} so, in order to construct conserved
quantities, it is necessary to have a Killing vector at hand to
construct the current $\jota a = \tc ab \xi_b$.

The \tb ab as defined in (\ref{tb}) is the one that arises
naturally from Noether's theorem, since  \eq{Noether} shows that
if \st\ admits a Killing vector we obtain from \tb ab a conserved
current \jota a.
 Thus,
for instance,  the $n(n+1)/2$ currents in Minkowski \st, are
obtained from \tb ab by contracting it with the corresponding
Killing vectors.

 The canonical energy-momentum tensor
\tc ab\   is not symmetric except for scalar fields. It is not
even   gauge invariant for gauge theories. Of course, in flat
space-time, it holds $\cd a \tc ab=0$. But, as we mentioned above,
it is worthwhile noticing that this is not even true for  curved
\st. Since, taking into account that the Lagrangian $\lag$ depends
on the coordinates only through the tensor fields $ \cd a \ten,\
\ten$ and $g_{ab}$, we can compute \beqa\label{707} \cd b (\lag) =
\dlag { \cd a \ten} \cd b \cd a \ten + \dlag {\ten} \cd b \ten \nn
= \dlag { \cd a \ten} \cd a \cd b \ten + \cd a \left( \dlag { \cd
a \ten}\right) \cd b \ten \nn - \dlag { \cd a \ten}\conm a b \ten\
\nn = \cd a \left( \dlag { \cd a \ten} \cd b  \ten \right) - \dlag
{ \cd a \ten}\ R^d_{\ cab}\ \til\ten{}^c_{\ d}\  , \eeqa where we
have used the field equations \eq{eqm},  and \eq{a5}. So, we get
\beq \cd a \tc ab = \dlag { \cd a \ten}\ R^b_{\ adc}\
\til\ten{}^{cd}\ . \eeq Thus, except for scalar fields,  \tc ab\
is not ``conserved"
 when space-time is curved.

Moreover, even in flat \st, for a Killing  field $\xi_b$\ it holds
$\cd a (\tc ab \xi_b)= \tc {[a}{b]} \cd a \xi_b$, so it vanishes
only for parallel $\xi_b$'s for general tensor fields, since
  \tc ab is not  symmetric. Then
we get  from \tc ab only $n$ conserved currents associated to the
parallel Killing vectors (translations). A similar result holds
for curved \st , even though $\cd a \tc ab \neq 0$. In fact, if
there exists a parallel Killing vector ($\cd a \xi^b=0$), \eq{qq}
shows that $\cd a (\tc ab \xi_b)= 0$.

Therefore, the  canonical energy-momentum tensor \tc ab\ is rather
an exception that occurs only when space-time admits parallel
Killing vectors. Our computations clearly show that, in general,
it is \tb ab\ and not \tc ab the one that arises naturally from
Noether's Theorem, so there is no reason to expect  much from \tc
ab . So, we find no reason to start from \tc ab and then
symmetrize it in order to get the right tensor
 $\tb ab$ (see for example
\cite{land,jack}). After all, we can always find a nonsense
correction to a wrong result to get the right one.

Notice that, $\tb ab =\tm ab$ means that for any scalar Lagrangian
depending on the tensor fields $ \cd a \ten,\ \ten$ and $g_{ab}$,
for fields satisfying the field equations, it must hold
\beqa\label{ee} 2\ \dlag {\!g_{ab}} = \cd c \left( \dlag { \cd c
\ten}\ \til{\ten}{}^{ab}\right) -\dlag {\cd a \ten}\ \nabla^b
\ten\ .\eeqa It is worthwhile noticing that \eq{ee} is a
consequence of $\til \lag =0$, since for any scalar  \lag\ we have
\beqa \til \lag\ {}^{ab}=0=\dlag { \cd c \ten}\ \til{\cd c
\ten}{}^{ab} +\dlag {\ten}\ \til{\ten}{}^{ab} + \dlag {g_{cd}}\
\til {g_{cd}}\ {}^{ab}, \eeqa taking into account that $\til
{g_{cd}}\ {}^{ab}=-\delta^a_c\delta^b_d-\delta^a_d\delta^b_c$, we
get \beqa 2\ \dlag {\!g_{ab}}\  = \dlag { \cd c \ten}\ \til{\cd c
\ten}{}^{ab} +\dlag {\ten}\ \til{\ten}{}^{ab} \ . \eeqa But, from
\eq{a8}, we have \beqa \til{\cd c \ten}{}^{ab}= \cd c \til \ten
{}^{ab} - \delta^a_c\nabla^b \ten\ . \eeqa Thus \beqa\ 2\ \dlag
{\!g_{ab}} = \dlag { \cd c \ten}\ \cd c \til \ten {}^{ab} +\dlag
{\ten}\ \til{\ten}{}^{ab} -\dlag {\cd a \ten}\ \nabla^b \ten\
.\eeqa Now, for fields satisfying the field equations \eq{eqm}, we
get \eq{ee}.

For instance, for a scalar field $\phi$ the first term in \eq{ee}
vanishes, since $\til\phi=0$, and so we get \beqa\ 2\ \dlag
{\!g_{ab}} = -\dlag {\pd a \phi}\ \partial^b \phi\ .\eeqa For
electromagnetic fields, we have $\til{A_d}{}^{ab}=-\delta^a_d
A^b$, and so \beqa 2\ \dlag {\!g_{ab}} = -\cd c \left( \dlag { \cd
c A_d}\ \delta^a_d A^b\right) -\dlag {\cd a A_c}\ \nabla^b A_c = -
\dlag { \cd a A_c} F^b_{\ c}\ .\eeqa Moreover, as the right
hand-side of \eq{ee} is a  symmetric tensor field, so  is the left
hand-side. Hence, for fields satisfying the field equation, we
have \beqa \cd c \left( \dlag { \cd c \ten}\
\til{\ten}{}^{ba}\right) -\dlag {\cd b \ten}\ \nabla^a \ten =\nn
\cd c \left( \dlag { \cd c \ten}\ \til{\ten}{}^{ab}\right) -\dlag
{\cd a \ten}\ \nabla^b \ten\ .\eeqa

Usually the metric energy-momentum tensor is defined through the
variation of the action \eq{action} (see for instance \cite{haw})
\beq\label{TM} \delta S:= \frac{1}{2}\int_\Omega \tm ab\ \delta
g_{ab}\ \dv\ , \eeq where $\delta g_{ab}$ ar arbitrary variations
of the metric vanishing everywhere except in  the
  interior of $\Omega$.
We can easily show that it coincides with the one defined in
\eq{tm} for, under the change $g_{ab}\rightarrow g_{ab}+\delta
g_{ab}$, \beqa \label{333} \delta \lag= \dlag {\cd a \ten}\ \delta
\cd a \ten + \dlag {g_{ab}}\ \delta g_{ab}\ . \eeqa But, according
to \eq{pd} \beq \delta \cd a \ten = \delta \left(\pd a \ten +
\Gamma ^b_{\ ca}\ \til \ten {}^c_{\ b}\right) =\delta  \Gamma
^b_{\ ca}\ \til \ten {}^c_{\ b}\ . \eeq Thus, by using the well
known relation \beq \delta \Gamma^b_{\ ca}= \frac{1}{2}\
g^{bd}\left( \cd a \delta g_{dc} +\cd c \delta g_{ad}-\cd d \delta
g_{ac}\right)\ , \eeq we can write the first term in \eq{333} as
\beqa\label{48} \dlag {\cd a \ten}\ \delta \cd a \ten =
\frac{1}{2} \dlag {\cd a \ten}\ \til \ten {}^{cb}\ \left( \cd a
\delta g_{bc} +\cd c \delta g_{ab}-\cd b \delta g_{ac}\right)\nn =
\frac{1}{2} \left(\dlag{ \cd a \ten}\    \til \ten{}^{bc} +\ \th
abc \right)\cd a \delta g_{bc}\ . \eeqa Therefore, under the
change $g_{ab}\rightarrow g_{ab}+\delta g_{ab}$ \beqa\label{304}
\delta \left( \lag \sqrt{|g|}\right) = \frac{1}{2}\left(2\ \dlag
{g_{ab}}\ -\cd c \left(\dlag{ \cd c \ten}\    \til \ten{}^{ab} +
\th cab \right) + g^{ab} \lag \right)\delta g_{ab}\ \sqrt{|g|}\nn
+\frac{1}{2}\ \cd c \left( \left(\dlag{ \cd c \ten}\    \til
\ten{}^{ab} +  \th cab \right) \delta g_{ab}  \right)\ \sqrt{|g|}\
,\nn \eeqa where we have used the well known result \beq
 \delta  \sqrt{|g|}
=\frac{1}{2}\ \sqrt{|g|}\
 g^{ab}\ \delta g_{ab}\ .
\eeq
Finally, by integrating \eq{304} over any compact region $\Omega$,
taking arbitrary symmetric tensor fields $\delta g_{ab}$ vanishing everywhere except in  its  interior,
we show that definitions \eq{tm} and \eq{TM} coincide.

Equation \eq{48} shows that the term between brackets in \eq{tm}
arises from the Lagrangian dependence on the affine connection. In
particular, it is absent for scalar or electromagnetic fields.
Thus, in these cases, we have \beqa\label{tmel} \tm ab:= 2\ \dlag
{g_{ab}}\  + g^{ab} \lag\ . \eeqa In these cases, the ``tilde
calculus" turns out also to be unnecessary. In fact, there is a
simpler definition for the  energy-momentum tensor  for Maxwell's
Theory \cite{q}
\begin{eqnarray}
T_{E.M.}^{ab}:= -2\ \frac{\partial\lag}{\partial \F ac}\ F^b_{\
c}+ g^{ab} \lag \ ,
\end{eqnarray}
which turns out to be symmetric and gauge invariant, for any field
theory where the Lagrangian \lag\ is a local function of $\F a b$,
the exterior derivative $\partial_a A_b-\partial_b A_a$, of a
one-form field $A_b$.

\section{Conclusions}

Summarizing, we have shown that the Belinfante energy-momentum is
the one that arises naturally from Noether theorem when the metric
has isometries, and all the currents are written as $\jota a = \tc
ab \xi_b$.  Moreover, it coincides with \tm ab for general tensor
field theories.

On the other hand, the utility of our  definition of $\til {\bm
T}$ is apparent if we take into account that most of the equations
of this work contain at least one tilde.


\begin{acknowledgements}
I am grateful to Jorge Solomin
for valuable discussions.  This work
was supported in part by CONICET, Argentina.
\end{acknowledgements}


\end{document}